\documentclass{emulateapj}
\usepackage{psfig}

\def\ltsima{$\; \buildrel < \over \sim \;$}
\def\simlt{\lower.5ex\hbox{\ltsima}}
\def\gtsima{$\; \buildrel > \over \sim \;$}
\def\simgt{\lower.5ex\hbox{\gtsima}}

\def\gs{\mathrel{\raise0.35ex\hbox{$\scriptstyle >$}\kern-0.6em
\lower0.40ex\hbox{{$\scriptstyle \sim$}}}}
\def\ls{\mathrel{\raise0.35ex\hbox{$\scriptstyle <$}\kern-0.6em
\lower0.40ex\hbox{{$\scriptstyle \sim$}}}}

\def\andxii{AndXII}



\slugcomment{{\sc Accepted in ApJ, July issue: }  }

\begin{document}

\title{Strangers in the night: Discovery of a
dwarf spheroidal galaxy on its first Local Group infall}
\author{S.\ C.\ Chapman\altaffilmark{1,2,3}, 
J.\ Pe\~narrubia\altaffilmark{2}, 
R.\ Ibata\altaffilmark{4}, 
A.\ McConnachie\altaffilmark{2}, 
N.\ Martin\altaffilmark{5}, 
M.\ Irwin\altaffilmark{1}, 
A.\ Blain\altaffilmark{3},
G.~F.~Lewis\altaffilmark{6},   
B.\ Letarte\altaffilmark{3},
K.\ Lo\altaffilmark{7},
A.\ Ludlow\altaffilmark{2},   
K.\ O'neil\altaffilmark{7}
}
\altaffiltext{1}{Institute of Astronomy, Madingley Road, Cambridge, CB3 0HA, U.K.}
\altaffiltext{2}{University of Victoria, Victoria BC, V8P 1A1 Canada}
\altaffiltext{3}{California Institute of Technology, Pasadena, CA\,91125}
\altaffiltext{4}{
Observatoire de Strasbourg, 11, rue de l'Universit\'e, F-67000, Strasbourg, France} 
\altaffiltext{5}{Max-Planck-Institut fur Astronomie, Konigstuhl 17
69117 Heidelberg, Germany}
\altaffiltext{6}{
Institute of Astronomy, School of Physics, A29, University of Sydney, NSW
2006, Australia}
\altaffiltext{7}{NRAO,  P.O.\ Box 2, Green Bank,  WV 24944}

\begin{abstract}
We present spectroscopic observations of the \andxii\ dwarf spheroidal galaxy
using DEIMOS/Keck-II,
showing it to be moving rapidly through the Local Group 
(-556~km/s heliocentric velocity, -281~km/s relative to Andromeda),
falling into the Local Group from $\sim$115 kpc
beyond Andromeda's nucleus. 
\andxii\ therefore represents a dwarf galaxy plausibly falling 
into the Local Group
for the first time, and never having experienced a dense galactic
environment. From Green 
Bank Telescope observations, a limit on the H{\sc I} gas mass  of $<3\times10^3$~M$_\odot$
suggests that \andxii's gas could have been removed prior to experiencing the
tides of the Local Group galaxies.
Orbit models suggest the dwarf is close to the escape velocity of
M31 for published mass models. \andxii\ is  
our best direct evidence for the late infall of satellite galaxies,
a prediction of cosmological simulations.
\end{abstract}


\section{Introduction}
\label{txt:intro}

Dwarf galaxies and stellar streams represent the  visible remnants of the
merging process by which the halos of galaxies  are built up. They can be used
to unravel the hierarchical formation  of their host galaxies \citep{whiterees}.
%
Dwarf galaxies are also the systems with the highest mass-to-light ratio
found in the Universe (e.g., Mateo et al.\ 1998), 
which makes them one of the best
laboratories for investigating the nature of dark matter
(e.g., Pe\~narrubia et al.\ 2007). 

In the Cold Dark Matter paradigm a few Myrs after the Big Bang, the
distance between proto-dwarf galaxies and their hosts were much smaller. 
Early structure formation left dwarf galaxies over a large range
of distances from the host galaxy with small radial velocity components
(they were located approximately at their orbital apocenter). The largest dark
matter over-density in the local volume attracted these systems so that they
eventually merged moving on high eccentric orbits. In this picture, those
substructures initially lying close to the parent galaxy
were the first ones to be accreted, whereas those with large initial
separations either are accreting now or have yet to accrete. An important
prediction implicit in this scenario is that all accreting substructures are
bound to the host galaxy; only strong interactions, like three-body
encounters, might provide enough energy for them to escape the host systems.
Also interesting is the fact that the late accretion events are expected to
move on highly eccentric orbits, with large radial velocity components with
respect to the host. Moreover, these systems correspond to the objects that
formed furthest from the host galaxy, thus sampling a
satellite galaxy population that has mostly 
lived in a very different environment from the Local Group.
It is only now, after they have been
accreted, when strong gravitational fields start to alter their original
properties (mass, dark matter and distributions, gas fraction, etc.).
High-speed dwarf galaxies, moving with velocities close to the escape
velocity, therefore represent an interesting theoretical and observational
target that might provide important insights into the effects of dynamical
evolution on the properties of dwarf galaxies.

It is believed that the satellites of the Milky Way and M31 
have been accreted at relatively early times and their properties have been
molded by their interactions with their parent galaxies.
There will undoubtedly be
mergers in the future of 
satellite systems within the Local Group (e.g., the Magellanic Clouds
-- Kallivayalil, van der
Marel \& Alcock 2006; Pedreros, Costa \& Mendez 2006).
There is also the case of the rather extreme orbit of the
Leo~I dwarf spheroidal \citep{mateo}, which is
now moving rapidly outwards from the Local Group on the limit of being bound.
Recent observations \citep{sohn} support a picture where Leo~I has been
tidally disrupted on several perigalactic passages of a massive galaxy.
Here we report on kinematic observations of \andxii,
a dwarf spheroidal which is a candidate for first infall into the Local
Group (LG).

\section{Observations}

The dwarf spheroidal galaxy \andxii\ lies at 105~kpc (projected) from 
the nucleus of M31, the faintest dwarf
galaxy in the M31 outer halo MegaCam/CFHT survey data \citep{martin06,ibata07}, 
with an overdensity of  Red Giant Branch (RGB) stars with similar metallicity
(Fig.~1).
\andxii\ was followed
up with imaging from a SUPRIMECam/Subaru survey of M31 halo 
substructures (Martin et al.\ in prep) to 
improve the distance estimate to \andxii\ and better understand its 
stellar populations.
Spectra for candidate \andxii\ stars were obtained  with the DEIMOS spectrograph on the Keck-II telescope (Figs.~1 \& 2).
Multi-object Keck observations
with DEIMOS \citep{deep2} were made on 2006 Sept 21--24,
in photometric conditions and excellent seeing of 0.6\arcsec. 
We used the 600~line/mm grating, achieving resolutions $\sim3.5$\AA\ and
probing the observed wavelength range from 0.56--0.98~$\mu$m.
Exposure time was 140 min, split into 20-min integrations. Data
reduction followed standard techniques using the DEIMOS-DEEP2
pipeline \citep{deep2},
debiassing, flat-fielding, extracting, wavelength-calibrating and
sky-subtracting the spectra.

The  radial velocities of the stars were  
then measured with respect to a Gaussian
model of the Calcium-II triplet (CaT) absorption lines \citep{wilkinson}.
By fitting the  three strong CaT lines separately
(Fig.~2), an  estimate of  the radial
velocity accuracy  was obtained  
(Table~1), with typical
uncertainties of  $5$\ km/s to $12$\ km/s, before accounting for systematic
errors from sky lines (which can add another $5-10$\ km/s).
Close to the  center of \andxii\
there is a clear  kinematic grouping of eight stars with cross-correlation
peaks greater than 0.1 at a velocity of
$\sim$-556~km/s heliocentric which also lie on the \andxii\ RGB (Fig.~1).
This corresponds to -281~km/s relative to M31 after removing Milky Way motions.
Although we targeted 49 stars in the vicinity of the \andxii\ overdensity,
no other stars are associated to this dwarf by their kinematics;
all other stars lie more than $5\sigma$ away from the kinematic
grouping of stars associated with the \andxii\ CMD. The \andxii\ 
stars have a 
velocity dispersion of 5~km/s before accounting for instrumental errors,
however the velocity errors are sufficiently large
that a meaningful limit on the mass of \andxii\ is difficult to constrain.
An average metallicity of [Fe/H]=-1.9$\pm$0.2 is estimated for \andxii\ 
from the equivalent widths of the CaT lines of the combined spectrum in
Fig.~2, adopting the technique described in Ibata et al.\ (2005).
However, the 5 highest S/N spectra have a more metal-rich median, 
[Fe/H]=-1.7, closer to the photometric [Fe/H]=-1.5 (Martin et al.\ 2006).
 
We also searched for neutral hydrogen at rest frequency 1420.406~MHz, 
shifted to the systemic velocity of \andxii.
Using the Green Bank Telescope (Nov.~29, 2006)
and the 12.5~MHz bandwidth of the spectrometer,
we integrated on \andxii\ using interspersed on/off observations totalling 1\,hr each.
The 9.8\arcmin\ beam size implies that the full stellar extent of \andxii\ is
well within the beam.
There is no detection down to 2.53\,mJy RMS, suggesting a limit on the H{\sc I}
gas mass of $<3\times10^{3}$~M$_\odot$ for an assumed linewidth of 4~km/s
(comparable to the raw dispersion of the RGB stars).


\section{Results}

Our new data measure the radial motion and constrain the distance to \andxii.
%
Since our spectroscopy tells us which amongst the candidates
are definite \andxii\ RGB stars, we can assess the distance directly.
We first fit isochrones from the library of Girardi et al.\ (2004) to the 
RGB, assuming the oldest age in the library (log(age)= 10.25); 
younger ages make the fits worse, and the implied
radial distance larger. 
Template isochrones were first extinction corrected using the 
Schlegel, Finkbeiner \& Davis (1998) extinction maps, E(B-V)=0.11.
We then attempted to fit isochrones of [Fe/H]=-2.3,-1.7,-1.5 and -1.3, 
over a range in distance modulus (DM) from 
24.1 to 24.9 (M31 has a DM=24.47 and a 
$I_{0,\rm Vega, TRGB}$=20.54 -- McConnachie et al.\ 2005). 
The best fitting isochrones are overlaid in Fig.~1,
with [Fe/H]=-1.5, DM=24.5 (D$_{\rm hel}$=810~kpc) being the best overall fit
to the RGB and likely horizontal branch stars.
Fits to the RGB shape and horizontal branch region were worse with other 
metallicities. Further analysis of the stellar populations in \andxii\ will
be presented in Martin et al.\ (in prep).
To assess a likely error range on the distance, we then employed 
the ``Tip of the Red Giant Branch" (TRGB) technique
(e.g., \citealt{mcconnachie04}). 
An absolute upper limit to the radial distance can be obtained by
assuming the brightest RGB star ($I_{0,\rm Vega}$=20.95) is at the TRGB.
This implies D$_{max}=950$ kpc (Fig.~3).
To constrain the minimum distance, 
we must assess what the maximum possible offset could be from this 
brightest RGB star.
In the absence of knowledge of the luminosity function of
\andxii, we analyse the well populated RGB of
the dwarf spheroidal, AndII \citep{mcconnachie05}. 
Our Girardi et al.\ (2004) isochrone fit to the RGB suggests that our
8 confirmed members of \andxii\ lie within the top magnitude of the RGB. 
With 1000 random samples of 8 of the brightest stars from the top magnitude 
of the AndII RGB, we constrain a probability distribution for the true TRGB 
for \andxii\ (corresponding to offsets up to 0.5-mag)  
and thereby the likely range of distances (Fig.~3).
The median distance for \andxii\ is 830$^{+50}_{-50}$~kpc
(interquartile range), in excellent agreement with the simple isochrone fit
performed earlier, and
updates the derived parameters of \andxii:
M$_v$ becomes -6.9, and the half-light radius, r$_{hb}$,
becomes 137~pc (previously estimated as 125~pc).


Fig.~3 summarizes our model of orbits for \andxii. 
Integrating the orbit back in time necessitates that \andxii\ likely 
came from beyond the Local Group's virial radius. 
We adopt a recently published mass model for M31 \citep{geehan05},
dominated by a NFW \citep{nfw}
dark matter halo with total virial mass of 
$\sim1\times10^{12}$ M$_\odot$.  
An object traveling at -281\ km/s with respect to M31 at time zero
(today) is considered over all possible heliocentric distances in the
gravitational potential of M31.
\andxii\ likely reached distances close to
725~kpc beyond M31 in the previous 10\,Gyr
(a substantial part of the journey to the
neighboring galaxy concentration around M81), and is an excellent
candidate for a dwarf falling into the Local Group for the first time.

The high radial velocity of \andxii\ together with the large separation to
its apparent host, M31, suggest that \andxii's orbit might be highly
eccentric, close to the limit of being bound to M31 and the Local Group.
Figure~4 shows \andxii\ relative to the circular and escape velocities of
M31 in the Geehan et al.\ (2005) model. \andxii\ is the most extreme satellite of M31, even more so than recently discovered satellites
AndXIV (Majewski et al.\ 2007) and AndXI, AndXIII (Martin et al.\ 2006,
Chapman et al.\ 2007).
In a CDM framework,  subhalos cannot be
accreted exceeding the escape velocity unless they suffered strong 3-body
interactions. 
In this framework, \andxii\ 
places a lower limit on the mass of M31 slightly larger than
estimates from the present models (Fig.~4; Geehan et al.\ 2005).
There is also
an unknown proper (transverse) motion to add to the radial velocity
-- it is unlikely for \andxii\ to be approaching
head-on, and therefore might be falling into the Local Group  
faster than the limiting v$_{hel}=-556$~km/s.

\section{Discussion}

Theoretical considerations have suggested that even at the present day, 
galaxies should still be 
falling into the Local Group from surrounding overdense structures.
Benson (2005) measured the orbital parameters of
infalling substructures from N-body simulations.
There was little evidence for any mass dependence in the distribution of
orbital parameters, and so the results are expected to be applicable to \andxii.
The orbital parameters in the Benson (2005) simulations were
determined at the point where the substructure first entered the virial
radius of the larger halo.


The large velocity and likely distance behind M31 suggest that 
\andxii\ is falling into the Local Group for the first time. 
%
The most likely place where systems like \andxii\ form is in filaments.
Ludlow et al.\ (in prep) have looked explicitly at high velocity structures
falling into halos of similar total mass to the Local Group at late times.
Substructures which have yet to fall into the surrogate Local Group 
clearly lie within a filamentary morphology.

At late times (today), one observes subhalos in N-body cosmological
simulations with high velocities because a large fraction of them move
on parabolic orbits.
The ones that move outwards are those that accreted
many Gyrs ago and are now close to escaping, perhaps the situation
with Leo~I \citep{mateo} and AndXIV (Majewski et al.\ 2007).
The ones that move inwards, like \andxii, are those that were accreted
later.



If \andxii\ has not yet experienced the strong gravitational environment
of the Local Group, its H{\sc I} gas content could be an interesting
evolutionary constraint. Mayer et al.\ (2006, 2007) show that
assuming a low density of the hot Galactic corona consistent with
observational constraints, dwarfs with V$_{\rm peak}<$30km/s will be completely stripped of their gas content 
on orbits with pericenters of 50~kpc or less.
In these objects most of the gas is removed or becomes ionized at the
first pericenter passage, explaining the early truncation of the star formation
observed in Draco and Ursa Minor.
Galaxies on orbits with larger pericenters and/or falling into
the Local Group at late times (like \andxii) 
should retain significant amounts of the centrally concentrated gas. 
These dwarfs would continue to form stars over a longer period.
While the dark matter mass in \andxii\ is uncertain, it is plausibly $>10^6$
M$_\odot$, with an
initial gas mass of $>10^4$ M$_\odot$.
If \andxii\ has not yet passed through the potential of the Local Group,
and no longer has a sizeable gas supply ($<3\times10^3$~M$_\odot$ from
our GBT measurement) it
must have been stripped of its neutral H{\sc I} through other means,
perhaps some of it through its Pop~III stars.

While the precise direction of origin is uncertain because of the unknown
tangential velocity,
\andxii\ could have come roughly from the direction of the Sculptor group
and beyond towards the M81 group,
although it could not have traversed the distance from the M81 group
($\sim3.5$~Mpc \citealt{karachentsev02}) in the age of the Universe.
The discovery of \andxii\ presents the best piece of evidence to date for the late accretion of satellites 
and sets a new benchmark for testing the mass of Local Group galaxies and simulations of galaxy formation.

\medskip
\section*{acknowledgements}
SCC acknowledges a fellowship from the Canadian Space Agency and an NSERC
discovery grant.
JP acknowledge Julio Navarro for finanacial support.
Data presented herein were obtained using the W.\ M.\ Keck
Observatory, which is operated as a scientific partnership among
Caltech, the University of California and NASA. The Observatory was
made possible by the generous financial support of the W.\ M.\ Keck
Foundation.


\begin{table*}
\begin{center}
\caption{Positions and parameters of the
eight stars confirmed as associated to \andxii.}
\label{tableSat}
\begin{tabular}{llcccc}
\hline
{RA} & {dec} & vel (km/s) & vel-error & $g$-mag & $i$-mag \cr
\hline
  0 47 31.04 & 34 24 12.1 & -557.2 &  6.9 & 23.028 & 21.642 \cr
  0 47 24.69 & 34 22 23.9 & -565.8 & 10.1 & 23.129 & 21.658 \cr
  0 47 27.76 & 34 22  6.2 & -556.9 &  4.9 & 23.214 & 21.744 \cr
  0 47 28.63 & 34 22 43.1 & -561.0 &  5.1 & 23.382 & 21.974 \cr
  0 47 31.34 & 34 22 57.6 & -555.0 &  7.4 & 23.504 & 22.156 \cr
  0 47 26.65 & 34 23 22.8 & -548.1 & 10.5 & 23.880 & 22.542 \cr
  0 47 27.18 & 34 23 53.5 & -550.4 &  7.0 & 23.883 & 22.532 \cr
  0 47 30.60 & 34 24 20.3 & -559.5 & 11.2 & 23.903 & 22.657 \cr
\hline
\end{tabular}
\end{center}
\end{table*}

\eject
\centerline{\hbox{\psfig{file=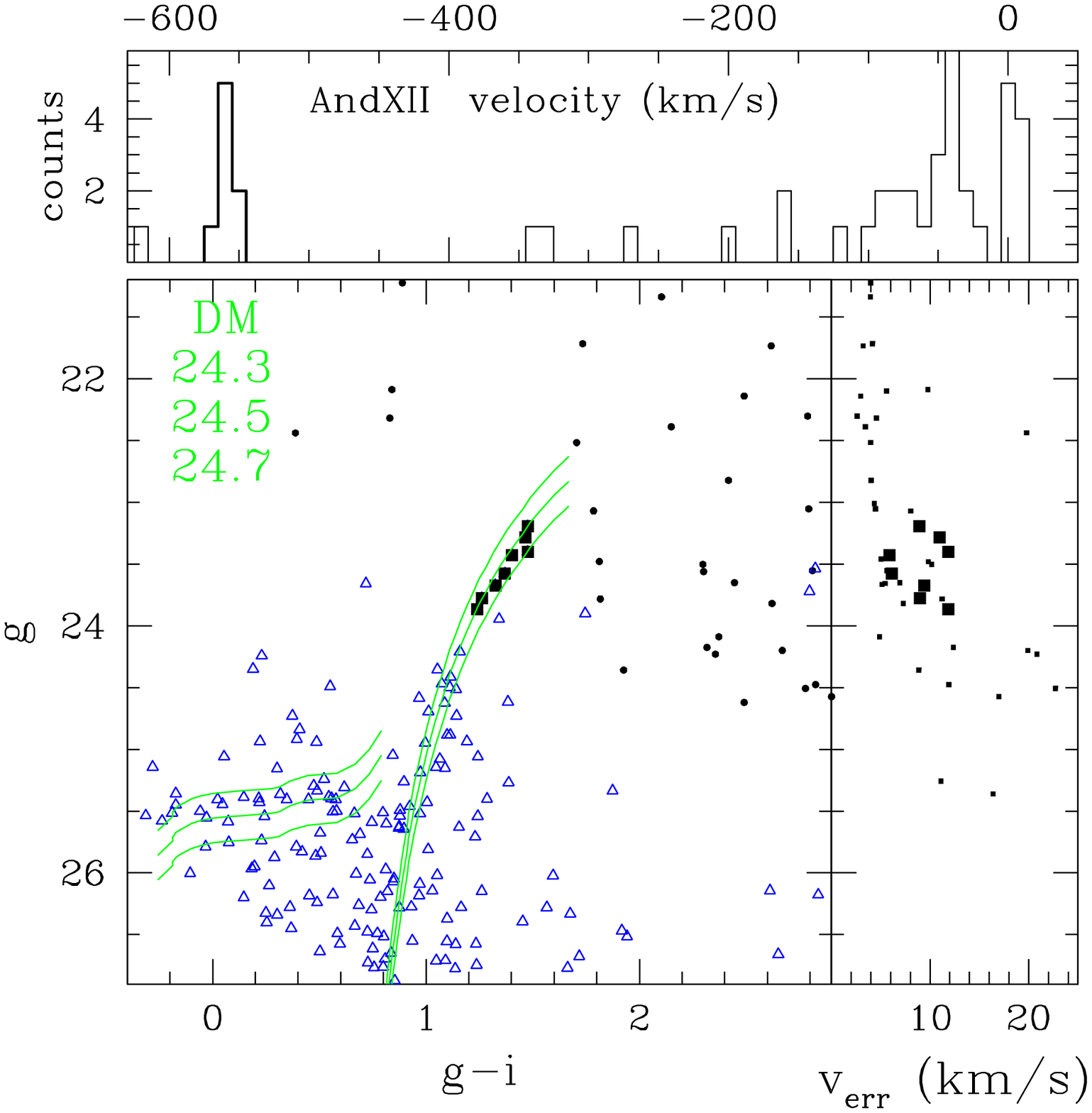,angle=0,width=3.5in}
\psfig{file=f1b.ps,angle=-90,width=3.5in}}}
\noindent
{\bf Figure 1.}
{\bf LEFT PANEL:}
On the top is shown the radial velocity histogram binned by 10~km/s.
The 8 stars in \andxii\ are bold. Stars at $>$150~km/s are likely
Galactic, while the remaining few stars are likely
members of the outer halo of M31.
On the left, the color-magnitude diagram of the field of \andxii\ with the 
8 stars in \andxii\ as large squares, the 49 stars with Keck radial velocities shown as squares, and all other stars within 3$'$ radius of
\andxii\ shown as open triangles. Stars selected for Keck spectroscopic
followup were taken from the CFHT-MegaCam image. All other stars
are shown from the Subaru SUPRIMECAM image.
Best fitting isochrones from the library of Girardi et al.\ (2004) are overlaid,
with [Fe/H]=-1.5, log(age)=10.25, and distance moduli of 24.3, 24.5, and 24.7.
On the right is shown the velocity errors for the
49 stars with radial velocity measurements.
{\bf RIGHT PANEL:} A representation of the dwarf galaxy
\andxii\ showing all stars with
photometric metallicity $-2.3<$[Fe/H]$<-1.3$ (Table~1 stars
highlighted, as well as the other 41 stars with DEIMOS velocities {\it not}
lying in \andxii) in the CFHT MegaCam M31 survey \citep{martin06,ibata07}.

\medskip 

\centerline{\hbox{\psfig{file=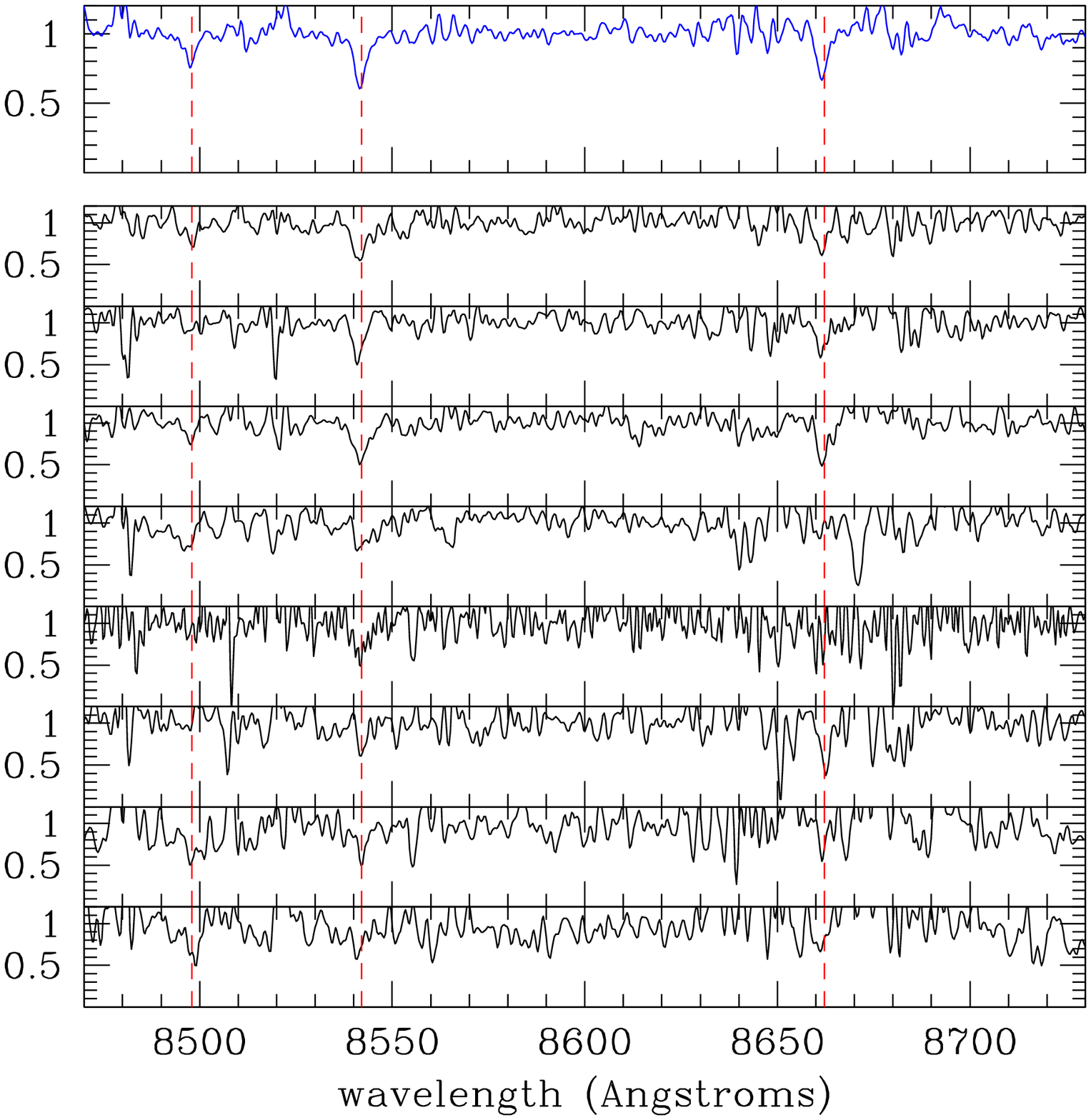,angle=0,width=5.5in}}}
\noindent
{\bf Figure 2.} The eight individual spectra, and the combined spectrum 
of stars from Table~1
(smoothed to the instrumental resolution)
spanning the $i$ = 21.6--22.6 range.
The CaT lines are shown in the restframe, with spectra shifted to zero
velocity.
\medskip

\centerline{\hbox{\psfig{file=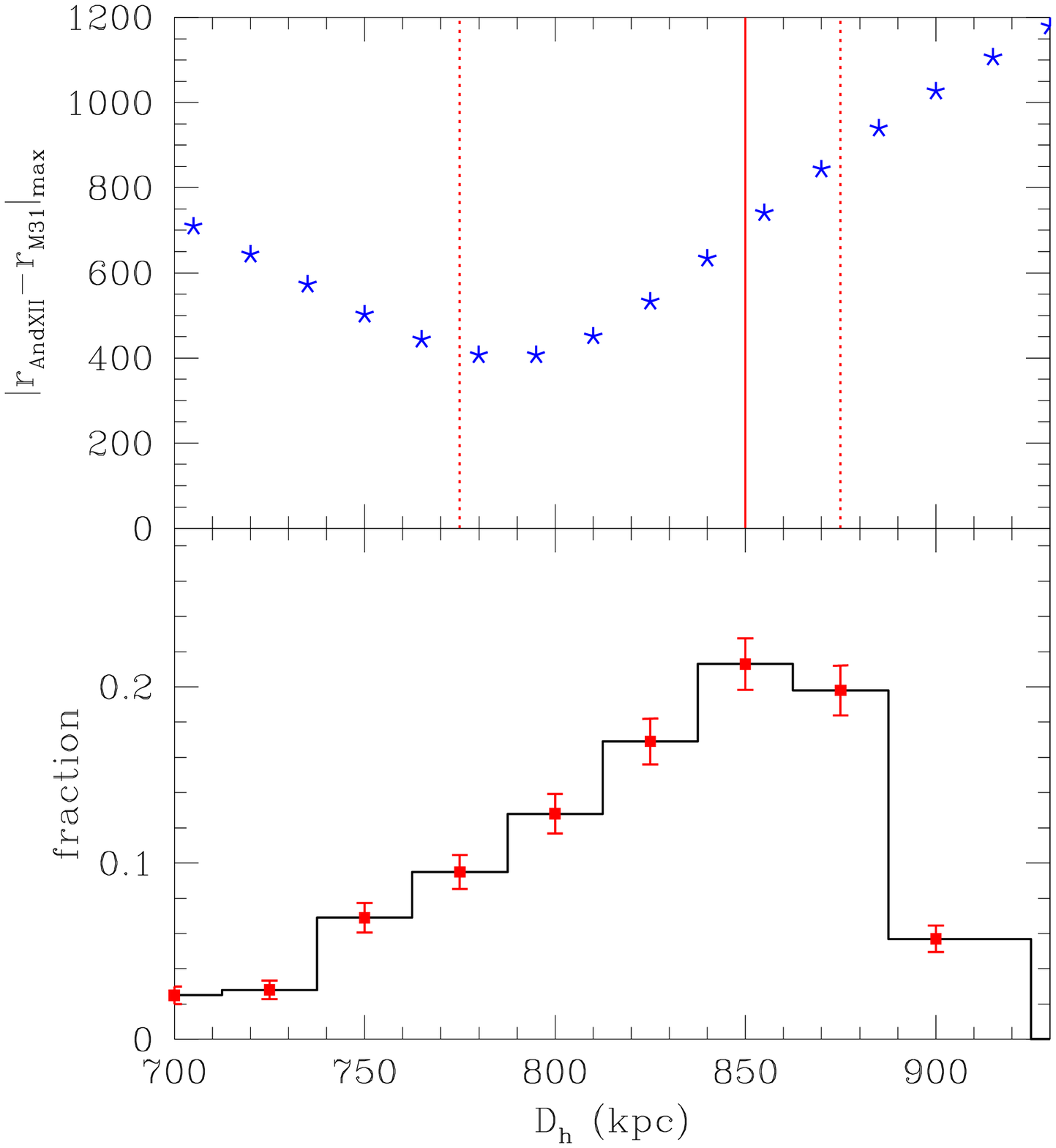,angle=0,width=6.0in}}}
\noindent{\bf Figure 3.}
{\bf Top panel:}
The distance \andxii\ would have reached from M31 in the last
10 Gyr, 725~kpc for the most probable radial distance from the
TRGB resampling distribution.
{\bf Bottom panel:}
The heliocentric distance probability curve for
\andxii, derived using our TRGB analysis and calibrated to an isochrone
fit to the RGB and HB.  
\medskip

\centerline{\hbox{\psfig{file=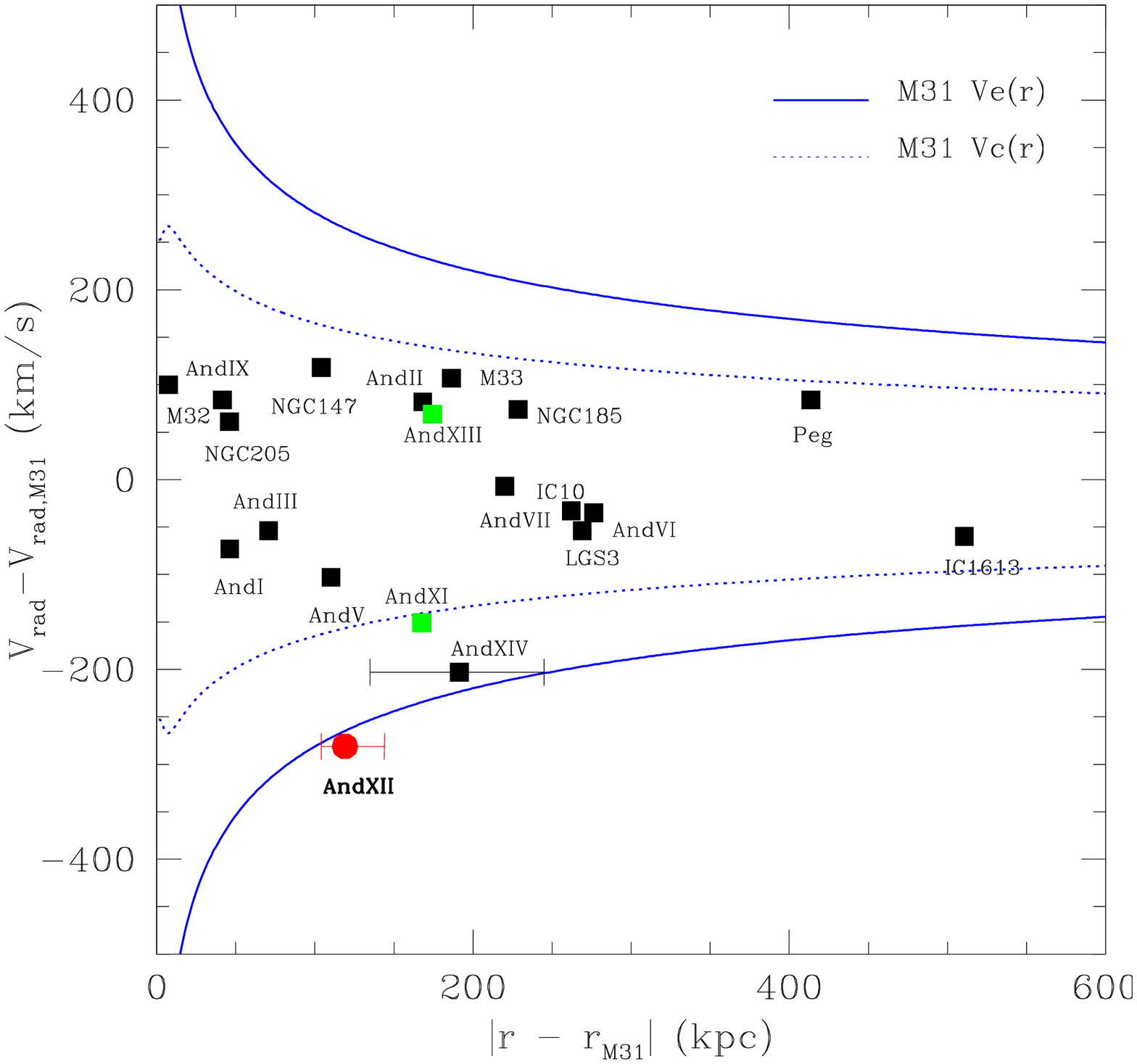,angle=0,width=6.0in}}}
\noindent{\bf Figure 4.}
The escape velocity is shown for M31 assuming the Geehan et al.\ (2005) mass model (solid line), derived from the circular velocity (dashed line),
as a function of three dimensional distance of M31 from its satellites
(data from Cote et al.\ 2000, and McConnachie et al. 2005).
\andxii\ is shown with a circle.
Recently discovered satellites AndXIV (Majewski et al.\ 2007) and
AndXI, AndXIII (Martin et al.\ 2006, Chapman et al.\ 2007) are 
highlighted.

\end{document}